\documentclass[11pt]{iopart}
\usepackage{comment}
\expandafter\let\csname equation*\endcsname\relax
\expandafter\let\csname endequation*\endcsname\relax
\usepackage{amsmath}
\usepackage{amssymb}
\usepackage{color}
\usepackage{graphicx}
\usepackage[sort&compress,numbers]{natbib}
\begin{document}
\date{\today}

\title[Quantum dissipative dynamics of a bistable system]{Quantum dissipative dynamics of a bistable system in the sub-Ohmic to super-Ohmic regime}

\author{Luca Magazz\`u$^{1,2}$, Angelo Carollo$^1$, Bernardo Spagnolo $^{1,2,3}$, Davide Valenti$^{1}$}
\address{$^1$ Dipartimento di Fisica e Chimica, Group of Interdisciplinary Theoretical Physics, 
Universit\`{a} di Palermo and CNISM, Unit\`a di Palermo, Viale delle Scienze, Edificio 18, I-90128 Palermo, Italy}
\address{$^2$ Radiophysics Department, Lobachevsky State University of Nizhny Novgorod, Russia}
\address{$^3$ Istituto Nazionale di Fisica Nucleare, Sezione di Catania, Italy}
\ead{luca.magazzu@unipa.it, angelo.carollo@unipa.it, bernardo.spagnolo@unipa.it, davide.valenti@unipa.it}

\begin{abstract}
We investigate the quantum dynamics of a multilevel bistable system coupled to a bosonic heat bath beyond the perturbative regime. We consider different spectral densities of the bath, in the transition from sub-Ohmic to super-Ohmic dissipation, and different cutoff frequencies.  The study is carried out by using the real-time path integral approach of the Feynman-Vernon influence functional. We find that, in the crossover dynamical regime characterized by damped \emph{intrawell} oscillations and incoherent tunneling, the short time behavior and the time scales of the relaxation starting from a nonequilibrium initial condition depend nontrivially on the spectral properties of the heat bath.
\end{abstract}



\vspace{2pc}

\submitto{\JSTAT}

\maketitle

\section{Introduction}
\label{introduction}
Real quantum systems are always in contact with noisy environments causing dissipation and decoherence. Quantum dissipation by phononic or electromagnetic environments is described well by the celebrated Caldeira-Leggett model~\cite{Caldeira1983} in which a quantum particle, the \emph{open system}, is linearly coupled to a \emph{reservoir} of $N$ independent quantum harmonic oscillators. Even if the coupling with the individual oscillators is weak, the overall dissipative effect may be strong, especially for macroscopic systems such as superconducting quantum interference devices~\cite{Caldeira1981,Weiss2012}. In the thermodynamical limit $N\rightarrow\infty$ the reservoir is a bosonic \emph{heat bath} and is described by the spectral density function $J(\omega)$, usually assumed to be of the form $\omega^{s}$ with a high-frequency cut-off. The special case $s=1$ gives the so-called Ohmic dissipation. In this case the quantum Langevin equation for the position operator of the particle has a memoryless damping kernel (frequency independent friction) and, in the classical limit $\hbar\rightarrow 0$, the heat bath reduces to a classical white noise source~\cite{Weiss2012}.\\
\indent In this work we study the bistable dynamics of a quantum particle coupled to an environment of which we vary the spectral density. Bistable potentials are ubiquitous both in the classical and quantum context~\cite{Spagnolo2012} as they are used to study the passage between potential minima separated by an energy barriers that can be classically surmounted or crossed via quantum tunneling. The quantum regime is generally characterized by the presence of several energy levels below the potential barrier. Due to the strong nonlinearity of the potential, these levels are organized in \emph{tunneling doublets} with internal energy separation much smaller than the interdoublet separation.  As a consequence the tunneling dynamics occurs on time scales much larger than that of the oscillations internal to the well, whose frequency is given by the separation between the first and the second doublet.\\
\indent The usual way of describing the tunneling dynamics in the presence of dissipation is by means of the two-level system (TLS) approximation. Within this approximation the Hilbert space of the particle is truncated to that spanned by the first two energy eigenstates. Despite its simplicity, the \emph{spin-boson} model, resulting from the TLS truncation of the Caldeira-Leggett model, displays several nontrivial features such as the coherent-incoherent transition, as the coupling strength is increased, and a further transition to the  localized phase at stronger coupling, the occurrence of which depends on the spectral content of the environment~\cite{LeHur2008,Wang2008,Weiss2012}.\\
\indent However, when the temperature is of the same order of  magnitude of the separation between the first and the second energy doublet, the TLS approximation is questionable, as higher energy levels are non-negligibly involved in the dynamics, and a beyond-TLS description becomes then necessary~\cite{Cukier1990,Gatteschi2006,Chatterjee2010,Johnson2011}. Nevertheless, theoretical investigations have been mainly focused on the dissipative dynamics of the spin-boson model~\cite{Leggett1987, EggerPRB1994,NesiPRB2007,Nalbach2010,Nalbach2013,Kast2013,Bera2014}.\\
\indent Regardless of the assumptions on the open system's Hilbert space, the interaction with an environment beyond the weak coupling regime yields intricate time-nonlocal effects in the reduced dynamics. To deal with these intricacies, several numerical techniques based on different strategies  have been developed so far. These include the quasi-adiabatic propagator path integrals (QUAPI)~\cite{Makri1995}, variationally optimized master equation approach~\cite{McCutcheon2011}, numerical renormalization group techniques~\cite{Daley2004,Bulla2008,Prior2010}, Monte Carlo path integral evaluation~\cite{Egger2000}, methods based on the wave function propagation~\cite{Wang2003}, stochastic techniques~\cite{Stockburger2002}, hierarchy equations~\cite{Ishizaki2005} and hybrid stochastic hierarchy equations approaches~\cite{Moix2013}.\\
\indent In this work, we consider the dynamics of a dissipative bistable system, beyond the TLS approximation, in a temperature regime in which the presence of the second energy doublet cannot be neglected. By using a nonperturbative generalized master equation with approximated kernels, derived within the Feynman-Vernon influence functional approach~\cite{Feynman1963,Grifoni1996,ThorwartPRL2000}, we investigate a regime of dissipation which is outside of the validity of the perturbative Born-Markov master equation approach. We study the reduced dynamics by varying the exponent $s$ in the crossover from the sub-Ohmic ($s<1$) to the super-Ohmic ($s>1$) dissipation regime. We consider also the effects of changing the cutoff frequency $\omega_{c}$, i.e. the contribution of the high-frequency modes to the open dynamics. Recently, for a quantum dot modeled as a TLS interacting with bosonic and electronic environments at zero temperature, a study on the effects of varying the spectral density function from sub-Ohmic to super-Ohmic has been performed~\cite{Wilner2015}.\\
\section{Model}
\label{model}
\begin{figure}[htbp]
\begin{center}
\includegraphics[height=5.5cm,angle=0]{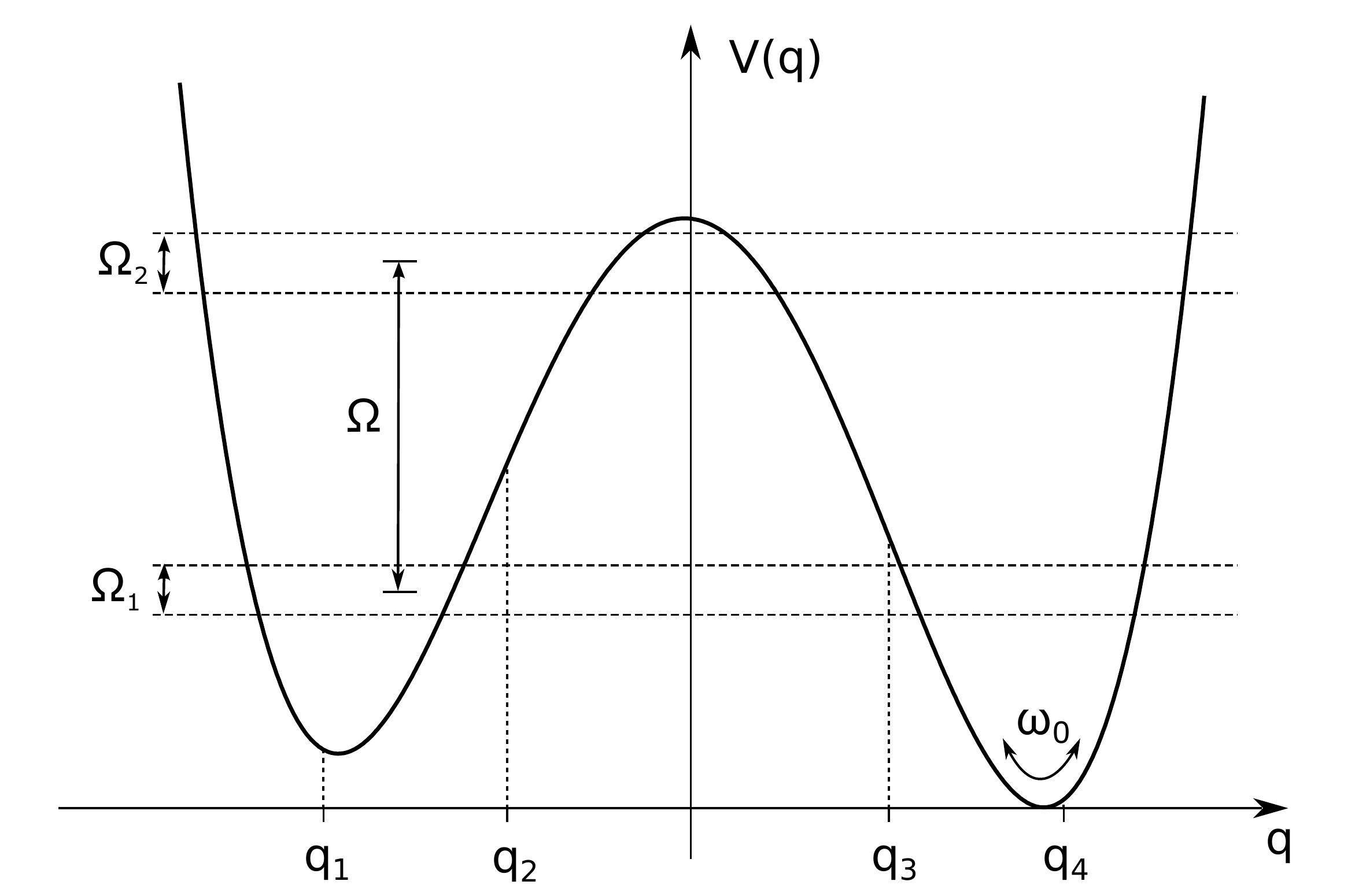}
\caption{\small{Potential $V_{0}$ with energy levels (horizontal lines) and DVR positions. 
The frequency $\omega_{0}$ is the oscillation frequency around the potential minima. 
The average inter-doublet frequency spacing, that is the characteristic frequency of the intrawell motion, is $\Omega=0.814~\omega_{0}$. The frequency spacings internal to the first and second energy doublet are $\Omega_{1}=0.123~\omega_{0}$ and  $\Omega_{2}=0.149~\omega_{0}$, respectively. These two frequencies characterize the tunneling dynamics of the bistable system.}}
 \label{fig1}
\end{center}
\end{figure}

\begin{figure}[ht]
\begin{center}
\includegraphics[height=5.4cm,angle=0]{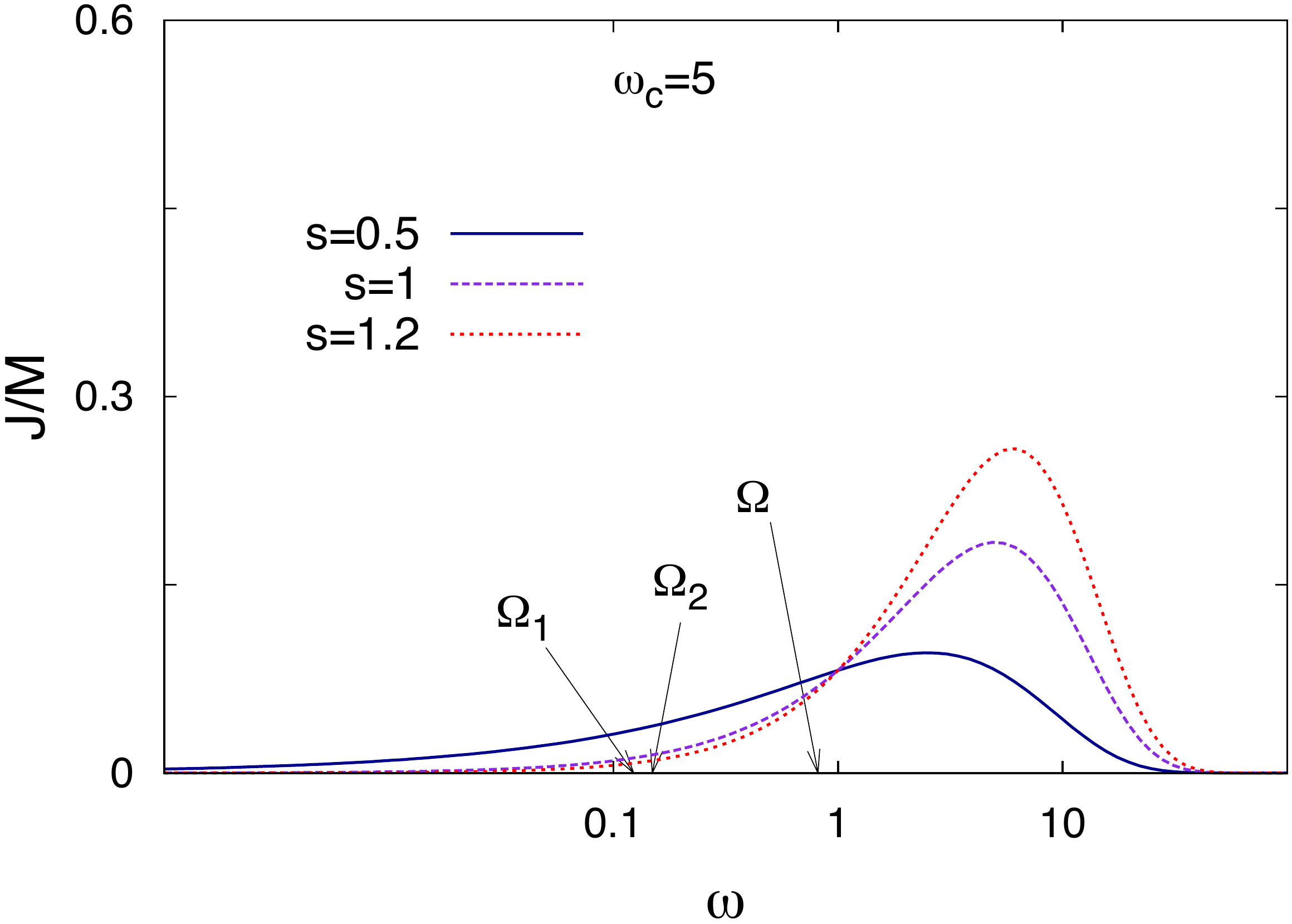}
\hspace{0.3cm}
\includegraphics[height=5.4cm,angle=0]{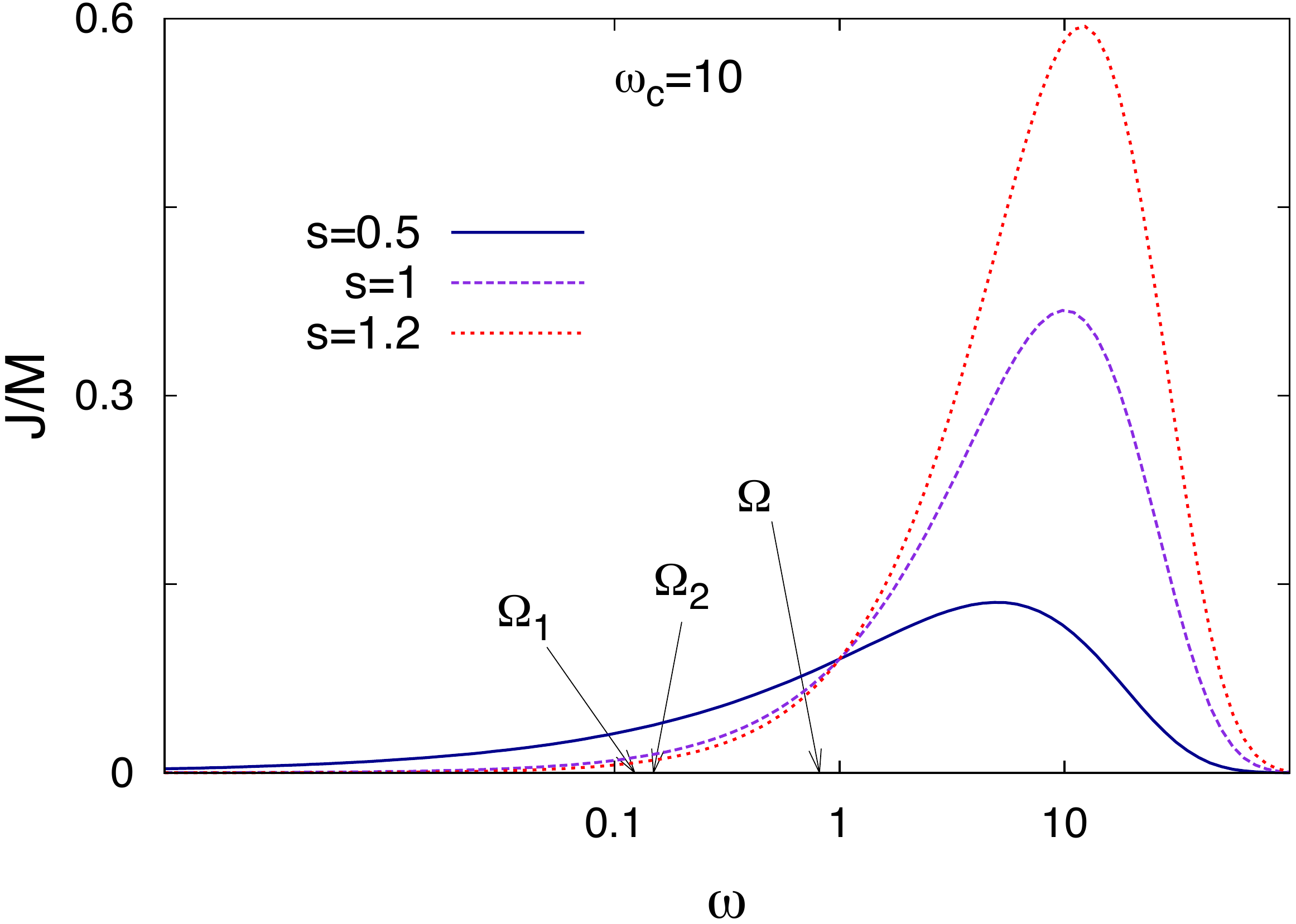}\\  
\caption{\small{Bath spectral density function (see Eq.~(\ref{J})) vs $\omega$. Three values of $s$ are considered for each of the two values of the cutoff frequency $\omega_{c}=5$ (left panel) and  $\omega_{c}=10$ (right panel). The characteristic frequencies of the isolated system, $\Omega$ and $\Omega_{i}$ (see Fig.~\ref{fig1}), are displayed for reference. Frequencies are in units of $\omega_{0}$}.}
\label{fig2}
\end{center}
\end{figure}

The open system $S$ is a quantum particle of mass $M$ and position operator $\hat{q}$ in the presence of a double well potential $V$. According to the Caldeira-Leggett model, the particle interacts linearly with a heat bath $B$ of $N$ independent quantum harmonic oscillators of positions $\hat{x}_{j}$ and momenta $\hat{p}_{j}$ . The full Hamiltonian of the model is 
\begin{equation}\label{H}
\hat{H}=\hat{H}_{S}+\hat{H}_{B}+\hat{H}_{SB},
\end{equation}
where $\hat{H}_{S}=\hat{p}^2/2M+V(\hat{q})$ is the open system Hamiltonian and 
\begin{equation}\label{H-SB}
\hat{H}_{B}+\hat{H}_{SB}=\frac{1}{2}\sum_{j=1}^N\left[\frac{\hat{p}_{j}^2}{m_{j}}+m_{j}\omega_{j}^2
\left(\hat{x}_{j}-\frac{c_{j}}{m_{j}\omega_{j}^2}\hat{q}\right)^2\right]
\end{equation}
is the free bath and system-bath (bilinear) interaction, whose strength is quantified by the set couplings $c_{j}$. The renormalization term quadratic in $\hat{q}$ is introduced to give a purely dissipative bath, i.e. a spatially homogeneous dissipation~\cite{Weiss2012}.\\
\indent The potential $V$ considered in this work is the double well potential with a couple of energy doublets under the barrier shown in Fig.~\ref{fig1}. 
Following the parametrization used in Ref.~\cite{ThorwartPRL2000}, $V$ is given by the quartic function  
\begin{equation}\label{V}
V(\hat{q})=\frac{M^2\omega^4_0}{64\Delta U}\hat{q}^4-\frac{M\omega^2_0}{4}\hat{q}^2-\epsilon\hat{q}.
\end{equation}  
The choice of the parameters of $V$ used in the present work is the same as that in Ref.~\cite{Magazzu2015}, where the multilevel dynamics in a sub-Ohmic environment with high frequency cutoff was investigated. Specifically we assume a static bias $\epsilon=0.02~\sqrt{M\hbar\omega_{0}^{3}}$ and a potential barrier (at zero bias) $\Delta U=1.4~\hbar\omega_{0}$. With our choice of parameters the first four energy levels are organized in two doublets separated by a frequency gap $\Omega\sim\omega_{0}$ with internal frequency separations $\Omega_{1},\Omega_{2}\ll\omega_{0}$ (see Fig.~\ref{fig1}).  \\ 
\indent This \emph{double-doublet} structure describes well that of the flux qubits used in the experiment on quantum annealing performed in Ref.~\cite{Johnson2011}.  In general, superconducting devices with externally tunable parameters~\cite{Poletto2009, Augello2010,Valenti2014} can produce theoretically and experimentally a variety of potential profiles and initial conditions. For such devices, typical values of $\omega_{0}$ are of the order of the GHz~\cite{Devoret1984,Devoret1985,Poletto2009}. In particular, the superconducting quantum devices of Ref.~\cite{Poletto2009} are the reference physical systems for our analysis. Moreover this kind of devices operate in a wide range of temperatures and coupling strengths~\cite{Han1991,Chiorescu2003,Chiarello2012}. The double-doublet system (DDS) is also used for modeling atomic Bose-Einstein condensates in double well potentials~\cite{Gillet2014} or proton transfer reactions~\cite{Cukier1990}.\\ 
\indent In the present work we assume the following algebraic spectral density function with exponential
cutoff 
\begin{equation}\label{J}
J(\omega)=M\gamma_{s}\left(\omega/\omega_{\text{ph}}\right)^{s-1}\omega e^{-\omega/\omega_{c}}.
\end{equation}
The bath is sub-Ohmic for $0<s<1$, Ohmic for $s=1$ and super-Ohmic for $s>1$. The parameter $\gamma_{s}$, with dimension of frequency, is the system-bath coupling strength. 
  The \emph{phonon frequency} $\omega_{\text{ph}}$ is a characteristic frequency of the reservoir,  introduced so that $J$ has the dimension of a mass times a frequency squared also  for $s\neq 1$. In what follows we set $\omega_{\text{ph}}=\omega_{0}$.\\
 \indent In Fig.~\ref{fig2} we show the spectral density functions $J(\omega)$ in the sub-Ohmic, Ohmic, and super-Ohmic regimes for two values of the cutoff frequency. There, the density of low frequency modes is the highest in the sub-Ohmic regime. On the other hand, the density of high frequency modes, especially at large cutoff frequency, is the largest in the super-Ohmic regime.\\
\indent We assume that the environment has a physical cutoff at $\omega_{c}$, which may be the Debye frequency of the medium in which the system is immersed. The values of the cutoff frequency $\omega_{c}$ used here are taken in the range $[5~\omega_{0},50~\omega_{0}]$. The bath modes with frequencies up to the frequency scale set by $\omega_{0}$ affect the particle dynamics through the quantum friction modeled by the Caldeira-Leggett Hamiltonian (Eq.~(\ref{H})) on the time scales of the intrawell motion or longer.  The modes with frequencies  $\omega\gtrsim\omega^{*}$ where  $\omega_{0}\ll\omega^{*}\sim\omega_{c}$, affect the system dynamics by renormalizing the mass according to  
\begin{equation}\label{M-star}
\Delta M_{\omega^{*}}\propto \int_{\omega^{*}}^{\infty}d\omega J(\omega)/\omega=M\gamma_{s}\omega_{c}^{s}\Gamma(s,\omega^{*}/\omega_{c}),
\end{equation}
where $\Gamma(z,x)=\int_{x}^{\infty} e^{-t}t^{z-1}dt$ is the \emph{incomplete gamma function}, related to the Euler Gamma function by $\Gamma(z,0)=\Gamma(z)$. These effects are fully taken into account by our approach without an explicit change in the mass. This can be shown by considering the quantum Langevin equation for the system, as done in Ref.~\cite{Weiss2012}. Of course, the frequencies higher than the cutoff frequency $\omega_{c}$ are suppressed by the exponential cutoff (Eq.~\ref{J}).\\
\indent   At fixed temperature $T$, the spectral density function $J$ determines the \emph{bath correlation function} defined by
\begin{equation}\label{L}
\begin{aligned}
L(t)=\frac{\hbar}{\pi}\int^{\infty}_0 d\omega J(\omega)\left( \coth{\frac{\hbar\omega\beta}{2}}\cos{\omega t}-i\sin{\omega t}\right),
\end{aligned}
\end{equation}
where $\beta=1/k_{B}T$.\\
\section{Feynman-Vernon influence functional approach}
\label{FV-approach}
\subsection{Exact path integral expression for the reduced density matrix}
\label{path.int.approach}
The reduced dynamics of the open system $S$ is given by the trace of the full density matrix $\rho_{SB}$ over the bath degrees of freedom  

\begin{equation}\label{RDM-a}
\rho(t)=Tr _{B}\left[U(t,t_{0})\rho_{SB}(t_{0})U^{\dag}(t,t_{0})\right].
\end{equation}
The time evolution operator is $U(t,t_{0})=\exp(-i\hat{H} (t-t_{0})/\hbar )$, with $\hat{H}$ the full Hamiltonian of the model given in Eq.~(\ref{H}). A factorized initial state of  the type $\rho_{SB}(t_{0})=\rho(t_{0})\otimes B^{\beta}$ is assumed, where $\rho(t_{0})$ is an arbitrary state of $S$ and $B^{\beta}=\exp(-\beta\hat{H}_{B})/Z$ is the thermal state of the bath.\\
\indent The reduced density matrix at time $t$ in the position representation has matrix elements $\rho_{qq'}=\langle q|\rho|q'\rangle$ given by
\begin{equation} \label{RDM}
\begin{aligned}
\rho_{qq'}(t)=\int dq_0\int dq'_0 G(q,q',t;q_0,q'_0,t_{0})\rho_{q_{0}q_{0}'}(t_{0}),
\end{aligned}
\end{equation}
where the propagator $G$ is the double path integral over the paths of the left and right coordinates $q$ and $q'$
\begin{equation}
\begin{aligned}\label{PIpropagator}
G(q,q',t;q_0,q'_0,t_{0})=\int^{q}_{q_0}\mathcal{D}q(\tau)\int^{q'}_{q'_0}\mathcal{D}^*q'(\tau) e^{\frac{i}{\hbar}(S[q(\tau)]-S[q'(\tau)])}\mathcal{F}_{FV}[q(\tau),q'(\tau)].
\end{aligned} 
\end{equation}
The functional $S[q(t)]=\int_{t_{0}}^{t}dt'\mathcal{L}_{S}(q(t'),t')$ is the action functional for the bare system. The Feynman-Vernon $\mathcal{F}_{FV}=\exp(-\Phi_{FV})$ derives from tracing over the bath degrees of freedom. The paths $q(\tau)$ and $q'(\tau)$ of the left and right coordinates are coupled in a time nonlocal fashion in the influence phase functional $\Phi_{FV}$, which, in terms of the combinations $x=q+q'$ and $y=q-q'$, takes on the following form
\begin{equation}\label{phiFV}
\begin{aligned}
\Phi_{FV}[y,x]=&\frac{1}{\hbar^{2}}\int_{t_0}^{t}dt'\int_{t_0}^{t'}dt''y(t') \left[L'(t'-t'')y(t'')+iL''(t'-t'')x(t'') \right]\\
&+i\frac{\lambda}{\hbar}\int_{t_0}^{t}dt' x(t')y(t').
\end{aligned}
\end{equation}
Here $L'$ and $L''$ are the real and imaginary part of bath correlation function defined in Eq.~(\ref{L}). The constant $\lambda=\int^{\infty}_0 d\omega J(\omega)/\omega =M\gamma_{s}\Gamma(s)\omega_{c}(\omega_{c}/\omega_{\text{ph}})^{s-1}$ is proportional to the so-called reorganization energy, which measures the overall system-bath coupling~\cite{Wang2008}.\\
 \subsection{Generalized master equation within the generalized noninteracting blip approximation}
\label{GME-DVR}
\begin{figure}[ht]
\begin{center}
\includegraphics[height=4.3cm,angle=0]{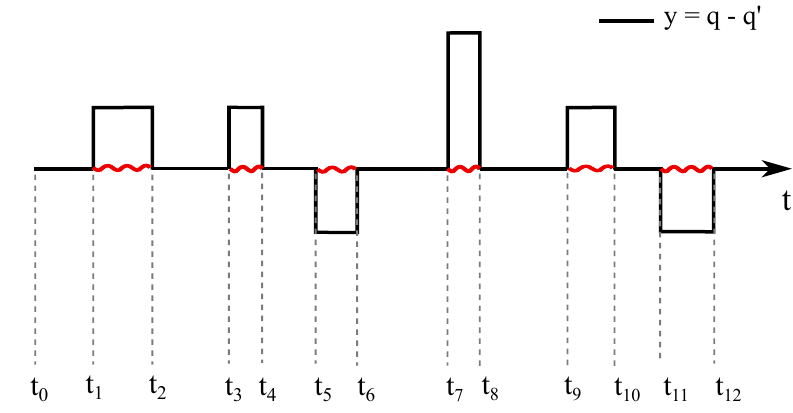}
\caption{\small{Example of a path of the reduced density matrix in the coordinate $y=q-q'$. The path consists of five intrawell blips and one tunneling blip between $t_{7}$ and $t_{8}$. The \emph{intrablip} interactions retained in the influence functional (see Eq.~(\ref{FV-DVR})) are denoted by the red wavy lines.}}
\label{fig3}
\end{center}
\end{figure}

\begin{figure}[ht]
\begin{center}
\includegraphics[height=5.3cm,angle=0]{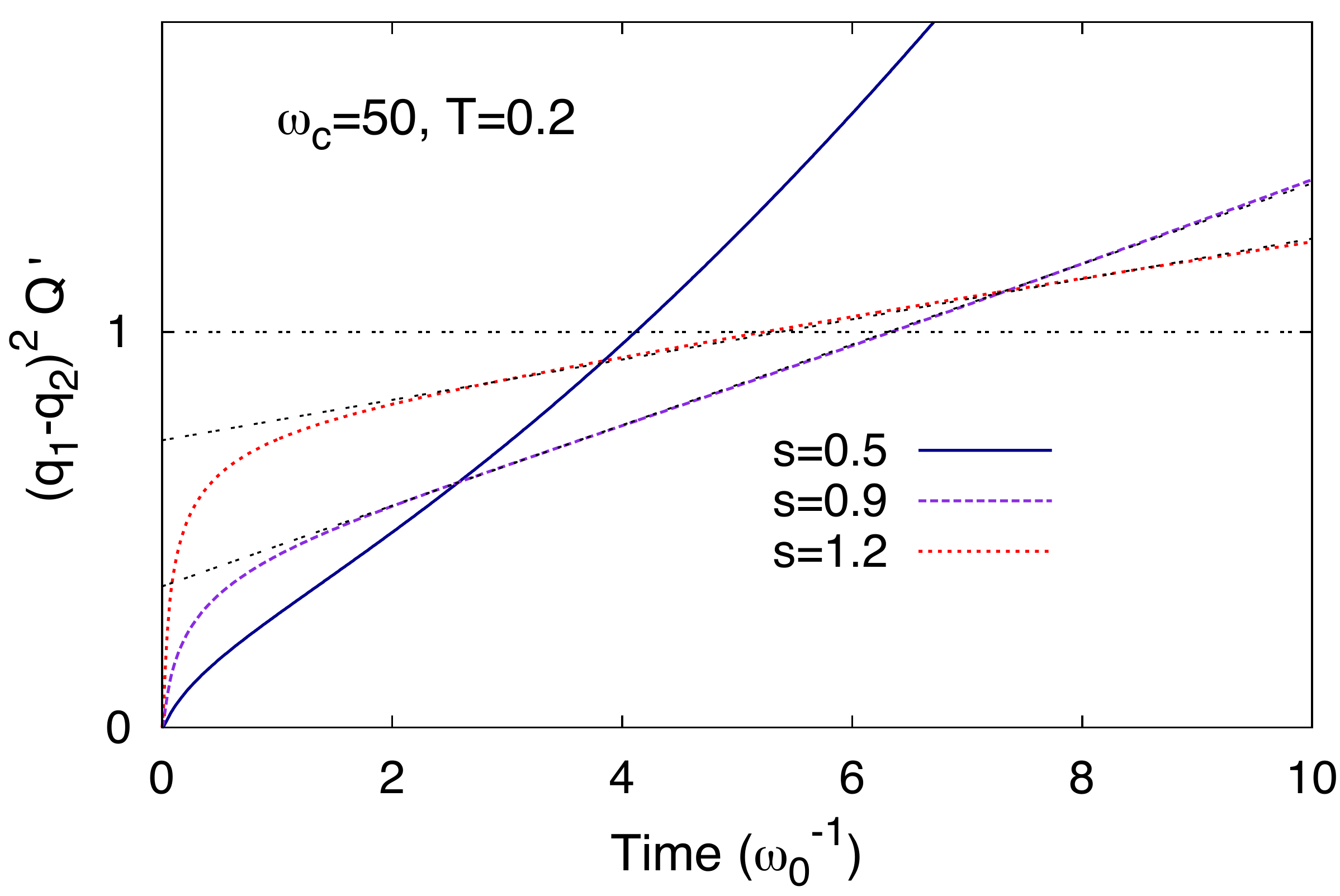}
\caption{\small{Real part of the pair interaction $Q$ (Eq.~(\ref{Q})) multiplied by the squared intrawell distance of the left well vs time. For $s\gtrsim 1$ the pair interaction assumes a linearized form on the time scales of the intrawell blips $\sim\omega_{0}^{-1}$. For $s<1$  $Q'$ grows rapidly so that the suppression of the interactions among different blips, operated by the real part of the influence phase (Eq.~(\ref{FV-DVR})), is effective on the same time scale. Temperature and cutoff frequency are in units of $\hbar\omega_{0}/k_{B}$ and $\omega_{0}$, respectively.}}
\label{fig4}
\end{center}
\end{figure}

\indent If the potential $V$ is harmonic, the propagator for the reduced density matrix can be evaluated analytically~\cite{Grabert1988} and one gets the exact dynamics of the dissipative harmonic oscillator. However, such an analytic solution does not exist for the nonlinear potential considered here. Nevertheless an approximate treatment is possible in a temperature regime where the system is not going to be excited to high energy levels and  the potential barrier is crossed by tunneling. In the present work this approximated treatment is based on the spatial discretization resulting from the truncation of the Hilbert space to that spanned by the first four energy eigenstates $|E_{1}\rangle,\dots,|E_{4}\rangle$. Performing on this restricted basis the unitary transformation $\bf{T}$, which diagonalizes the position operator $\hat{q}$ of matrix elements $\langle E_{i}|\hat{q}|E_{j}\rangle$, we pass to the discrete variable representation (DVR)~\cite{Harris1965,Light2000}. The DVR is given by the set of functions
\begin{equation}\label{DVR}
|q_{j}\rangle=\sum_{k=1}^{4}T^{*}_{jk}|E_{k}\rangle,
\end{equation}
\emph{focused} around the four position eigenvalues $q_{1},\dots,q_{4}$ depicted in Fig.~\ref{fig1}. The diagonal element $\rho_{ii}=\langle q_{i}|\rho|q_{i}\rangle$ of the reduced density matrix in the DVR, i.e. the population of the state $|q_{i}\rangle$, is the probability to find the particle in a region of space localized  around $q_{i}$ (see Fig.~\ref{fig1}).\\
\indent  In passing to the DVR, the exact reduced density matrix given in Eq.~(\ref{RDM}) assumes the form $\rho_{qq'}(t)=\sum_{q_{0},q'_{0}}G(q,q',t;q_{0},q'_{0},t_{0})\rho_{q_{0}q'_{0}}(t_{0})$, where the propagator has the formal expression 
\begin{equation}\label{G-DVR}
G(q,q',t;q_{0},q'_{0},t_{0})=\sum_{n=0}^{\infty}\int_{t_{0}}^{t}D_{n}\{t\}\mathcal{A}[q]\mathcal{A}^{*}[q']\exp(-\Phi[\xi,\chi]_{FV}).
\end{equation}
The sum is over the number $n$ of transitions of the paths and the symbol $\int_{t_0}^{t}D_{n}\{t_{j}\}$ denotes the sum $\sum_{\text{paths}_{n}}\int_{t_0}^{t}dt_{n}\int_{t_0}^{t_{n}}dt_{n-1}\dots\int_{t_0}^{t_2}dt_{1}$ over all path configurations with $n$ transitions at times $t_{j}$. In Eq.~(\ref{G-DVR}),
 the amplitude $\mathcal{A}[q]$ for the path $q(t_{k})$ of the isolated system includes the product of the transition amplitudes per unit time $\Delta_{ij}=\langle q_{i}|\hat{H}_{S}|q_{j}\rangle/\hbar$, relative to the transitions $|q_{i}\rangle\rightarrow|q_{j}\rangle$. The influence of the environment is encapsulated in the Feynman-Vernon influence phase, whose expression in the DVR is~\cite{Thorwart2001} 
\begin{equation}\label{FV-DVR}
\Phi[\xi,\chi]_{FV}=-\sum_{i=1}^{n}\sum_{j=0}^{i-1}\left[\xi_{i}Q'(t_{i}-t_{j})\xi_{j}+i\xi_{i}Q''(t_{i}-t_{j})\chi_{j}\right],
\end{equation}
where the so-called \emph{charges} at time $t_{k}$ are defined by $\xi_{k}=y_{k}-y_{k-1}$ and 
$\chi_{k}=x_{k}-x_{k-1}$.\\
The \emph{pair interaction} $Q(t)=Q'(t)+iQ''(t)$, which couples the $\xi$- and $\chi$-charges, is related to the bath force correlation function, defined in Eq.~(\ref{L}), by $L(t)=\hbar^{-2}d^{2}Q(t)/dt^{2}$. In terms of the parameter  $\lambda\equiv\lambda_{s}$ introduced in Eq.~(\ref{phiFV}), the pair interaction for a generic exponent $s$ reads~\cite{Weiss2012}
\begin{equation}\label{Q}
\begin{aligned}
Q(t)=\frac{\lambda_{s}}{\pi\hbar(s-1)}\left[1-(1+i\omega_{c}t)^{1-s}+\kappa^{s-1}f_{\kappa}(s-1,t)\right],
\end{aligned}
\end{equation}
where $\kappa=k_{B}T/\hbar\omega_{c}$ and $f_{\kappa}(s,t)=2\zeta(s,1+\kappa)-\zeta(s,1+\kappa+it/\hbar\beta)-\zeta(s,1+\kappa-it/\hbar\beta)$. The function $\zeta(z,q)=\sum_{n=0}^{\infty}1/(q+n)^{z}$ is 
 the Hurwitz zeta function, related to the Riemann zeta function by $\zeta(z,1)\equiv\zeta(z)$.\\
\indent  As a further approximation, in addition to the DVR, we restrict the sum over paths of the reduced density matrix in the propagator of Eq.~(\ref{G-DVR}) to the leading contributions. These are given by the class of paths consisting in sojourns in diagonal states interrupted by single off-diagonal excursions called \emph{blips}. In Fig.~\ref{fig3} is shown an example of path belonging to this class.\\
\indent Finally, in the dissipation regimes of intermediate to high temperature, on the scale fixed by $\hbar\omega_{0}$, considered here, the time nonlocal interactions among different blips in Eq.~(\ref{FV-DVR}), the \emph{inter-blip} interactions, can be neglected.  
This corresponds to a multilevel version~\cite{Grifoni1996, Thorwart2001} of the noninteracting blip approximation (NIBA)~\cite{Leggett1987,Weiss2012}. However, the relevant part of the interactions,  the intra-blip interactions indicated by the red wavy lines in Fig.~\ref{fig3}, are retained to all orders in the coupling strength. \\
\indent The NIBA is justified when, on a time scale comparable to the the average \emph{interblip} time distance, $Q'$ assumes its linear form with respect to time and $Q''$ becomes approximatively constant~\cite{Thorwart2001}. For intrawell blips this time is $\sim\omega_{0}^{-1}$ while for tunneling bilps
it is $\sim\Omega_{1,2}^{-1}\gg\omega_{0}^{-1}$.
 For $s\gtrsim 1$ the linearized form is reached on a time scale that is shorter at higher temperature~\cite{Weiss2012}. With the parameters  considered in this work, this time scale is that of the intrawell blips $\sim\omega_{0}^{-1}$ (see Fig.~\ref{fig4}). For $s<1$ the interblip interactions are suppressed because, as $Q'$ grows rapidly, the exponential cutoff due to the real part of the Feynman-Vernon influence functional becomes severe on the same time scale.  Therefore, the NIBA  is justified from the sub-Ohmic to the super-Ohmic regime for the intrawell motion. In turn this means that the same approximation scheme is valid \emph{a fortiori} for the tunneling blips, as their time scale is longer and their effective damping is larger due to the larger distance of DVR states separated by the potential barrier. The above reasoning justifies our generalized-NIBA treatment.\\
\indent Within the above approximations, from the exact path integral expression for $\rho$ it is possible to extract the following generalized master equation (GME) for the populations of the DVR states $|q_{i}\rangle$~\cite{Thorwart2001} 
\begin{equation}\label{GME}
\dot{\rho}_{ii}(t)=\sum_{j=1}^{4}\int_{t_{0}}^{t}dt'K_{ij}(t-t')\rho_{jj}(t')+I_{i}(t),
\end{equation}
where the inhomogeneous term $I_{i}$ vanishes at long times and is exactly zero if the system is prepared in one of the localized states $|q_{i}\rangle$~\cite{Thorwart2001}.  
In Eq.~(\ref{GME}) the populations are coupled to each other by means of the NIBA kernels 
\begin{equation}\label{K}
\begin{aligned}
K_{ij}(t)=&2\Delta_{ij}^{2}e^{-(q_{i}-q_{j})^{2}Q'(t)}\cos[\epsilon_{ij}t+(q_{i}-q_{j})^{2}Q''(t)] \qquad (i\neq j)\\
K_{jj}(t)=&-\sum_{l(\neq j)=1}^{4}K_{lj}(t),
\end{aligned}
\end{equation}
where $\epsilon_{ij}=(\langle q_{i}|\hat{H}_{S}|q_{i}\rangle-\langle q_{j}|\hat{H}_{S}|q_{j}\rangle)/\hbar$ is the generalization of the bias appearing in the TLS Hamiltonian in the left/right state representation. Note that, due to the prefactors $(q_{i}-q_{j})^{2}$ multiplying $Q$, the effective damping strength is much larger for  transitions between states in different wells. Moreover, the frequency scales $\Delta_{ij}$ associated to these transitions are smaller than those associated to the intrawell transitions. As a consequence,  by increasing the coupling strength $\gamma_{s}$, the tunneling oscillations are damped out already at relatively small values of the coupling, while the intrawell oscillations survive until much larger values are reached.\\ 
\indent The dynamical regime resulting from our choice of parameters is the \emph{crossover} regime of intrawell oscillations and incoherent tunneling, a regime which lies between the completely coherent and the fully incoherent dynamics. In Ref.~\cite{MagazzuPRE2015}, by using a beyond-NIBA scheme we have investigated this crossover dynamical regime in the Ohmic case down to temperatures for which NIBA-like approximations break down.\\
\indent Since the relaxation dynamics is governed by the incoherent tunneling and, as shown in Fig.~\ref{fig5}, the intrawell oscillations are damped out on relatively short time scales, a good estimate for the relaxation time, the time scale on which the system relaxes to equilibrium, is given by a Markov approximated version of Eq.~(\ref{GME})~\cite{ThorwartPRL2000,Thorwart2001}   
\begin{equation} 
\label{GME-markov}
 \dot{\rho}_{ii}(t)=\sum_{j=1}^{4}\Gamma_{ij}\rho_{jj}(t),  
 \end{equation}
where $\Gamma_{ij}=\int_{0}^{\infty}dt K_{ij}(t)$, with the kernels $K_{ij}$ given in Eq.~(\ref{K}). The solution for the population $\rho_{ii}$ is of the form $\rho_{ii}(t)=\rho_{ii}(\infty)+\sum_{k=1}^{3}c_{ik}\exp(-\Lambda_{k}t)$. The smallest of the rates $\Lambda_{k}$ gives the \emph{relaxation time}, defined as  $\Lambda_{\text{min}}^{-1}$. Notice that in this definition of relaxation time there is no reference to the initial condition. Eq.~(\ref{GME-markov}) does not capture transient oscillations and is accurate only in the fully incoherent regime. Nevertheless it gives a good estimate for the relaxation time also in the crossover dynamical regime~\cite{MagazzuPRE2015}. The master equation~(\ref{GME-markov}) has been used to obtain the dynamics and stationary populations in the presence of an external driving~\cite{Magazzu2013} and to address the problem of the escape from a quantum metastable state, starting from a nonequilibrium initial condition, with a strongly asymmetric bistable potential and Ohmic dissipation~\cite{Valenti2015}.\\ 

\section{Dynamics and relaxation times}
\label{results}
\begin{figure}[ht]
\begin{center}
\includegraphics[height=5.2cm,angle=0]{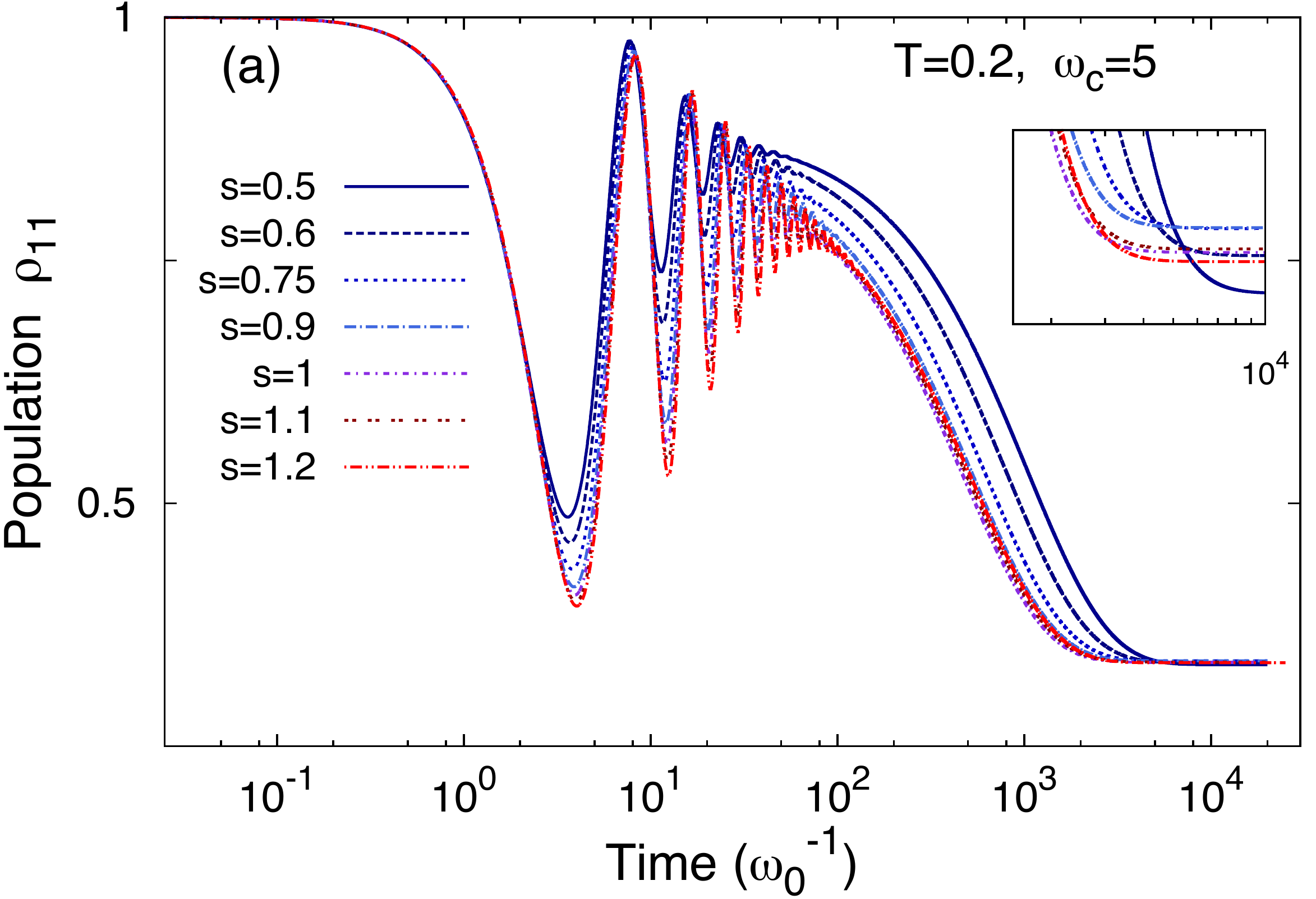}\hspace{0.2cm}
\includegraphics[height=5.2cm,angle=0]{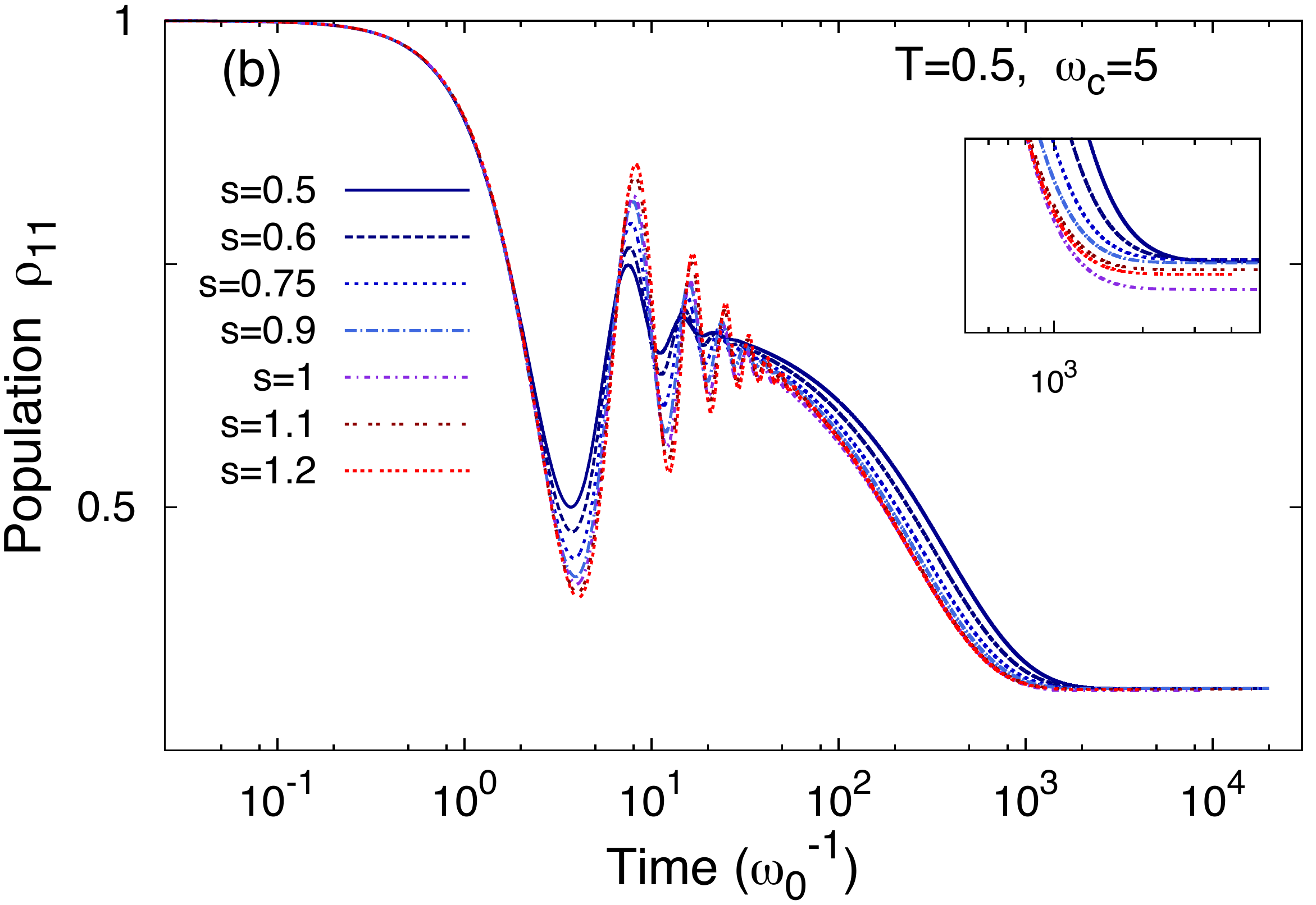}\\
\includegraphics[height=5.2cm,angle=0]{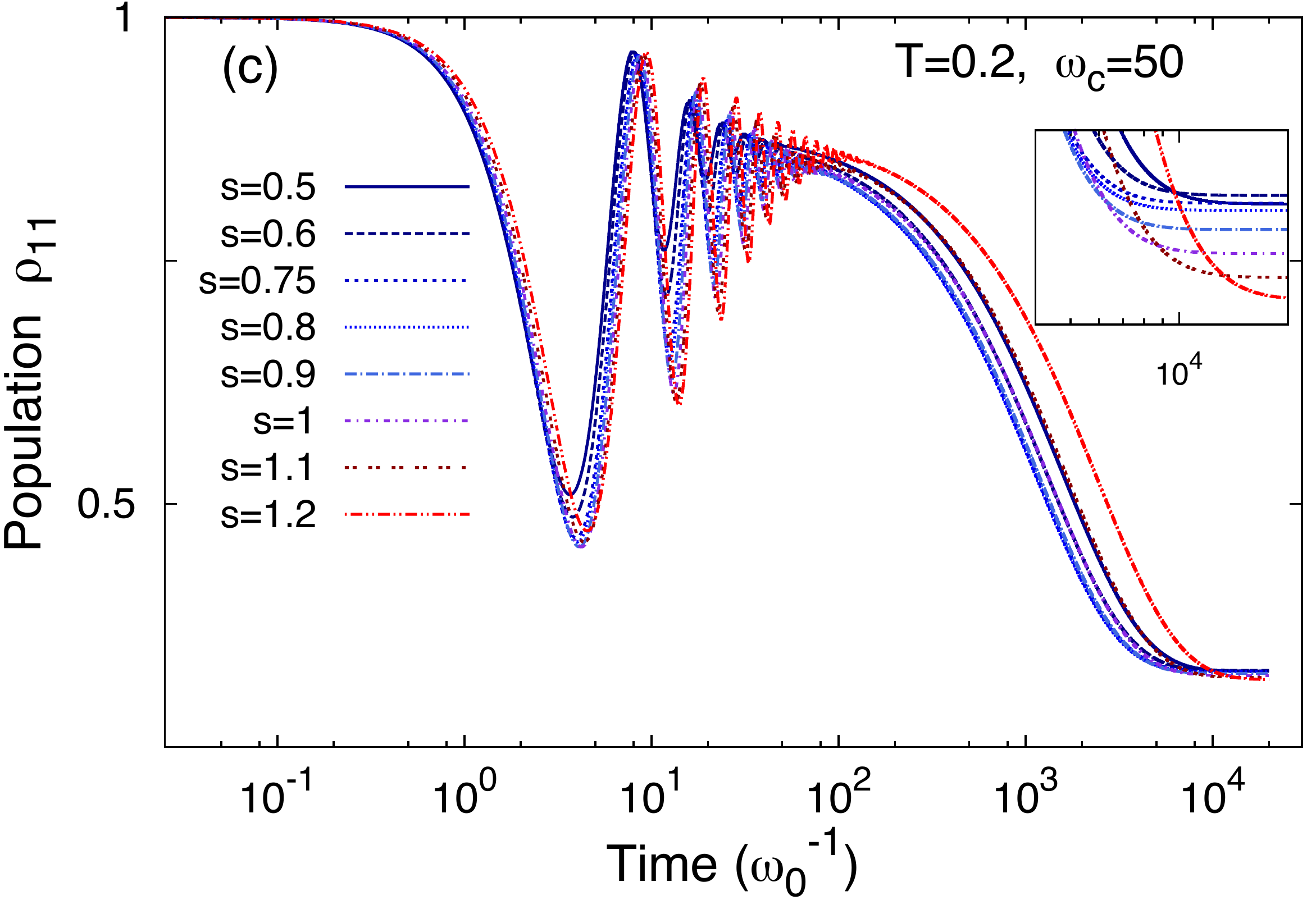}\hspace{0.2cm}
\includegraphics[height=5.2cm,angle=0]{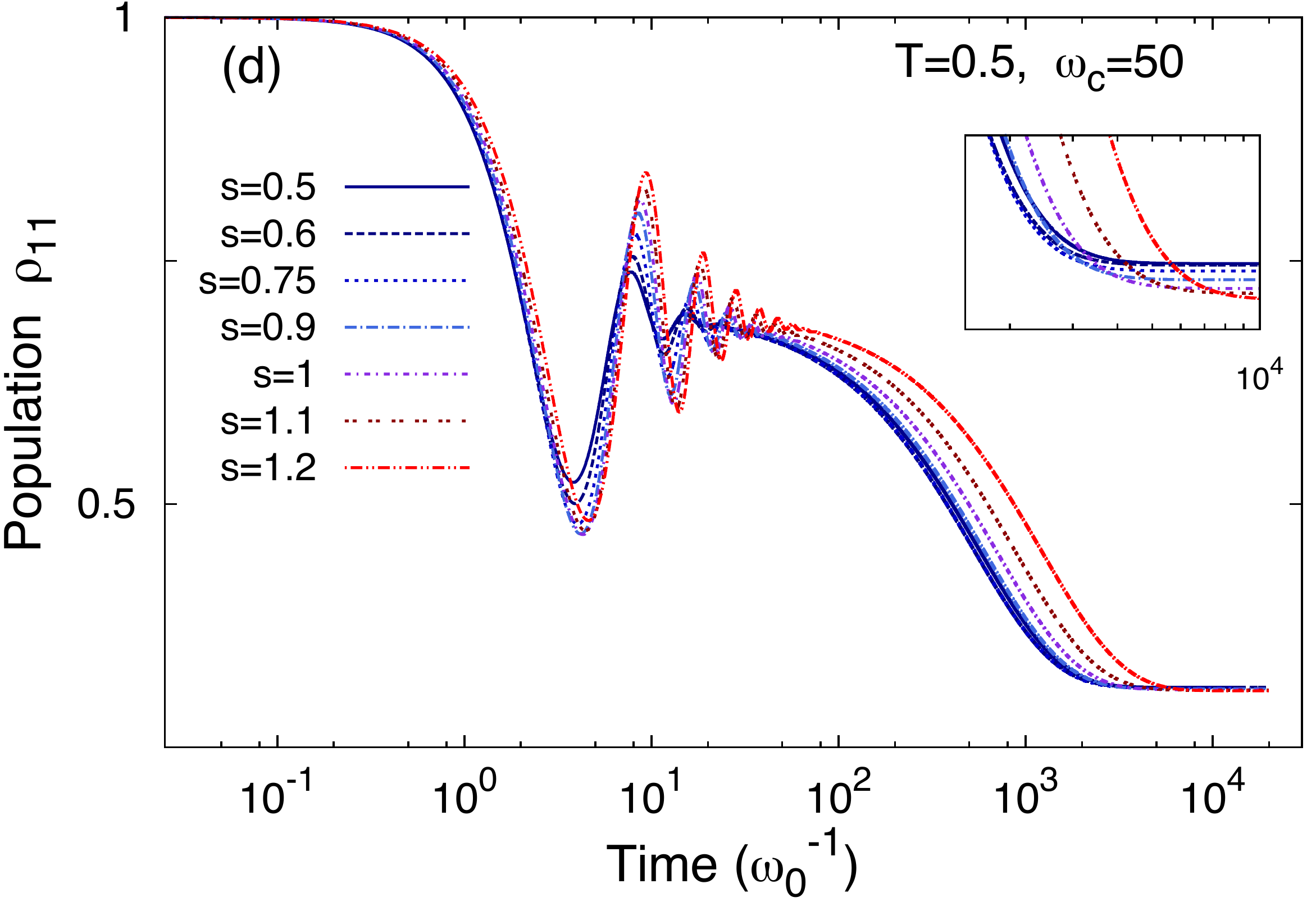}\\
\caption{\small{Time evolution of the population $\rho_{11}$ of the state $|q_{1}\rangle$ for different values of $s$, at two cutoff frequencies $\omega_{c}=5$ (upper panels) and $\omega_{c}=50$ (lower panels) and for two temperatures $T=0.2$ (left panels) and $T=0.5$ (right panels). The coupling strength $\gamma_{s}$ is fixed to the value $0.1~\omega_{0}$. Temperatures and frequencies are in units of $\hbar\omega_{0}/k_{B}$ and $\omega_{0}$, respectively.}} 
\label{fig5}
\end{center}
\end{figure}
In this section we show the results for the dynamics obtained from the generalized master equation~(\ref{GME}) with the NIBA kernels of Eq.~(\ref{K}). Calculations are performed by varying $s$, the exponent of $\omega$ in the spectral density function $J(\omega)$, in the range $0.5\leq s \leq 1.2$, and for different cutoff frequencies $\omega_{c}$ (see Eq.~(\ref{J})). We consider the two temperatures $T=0.2,~0.5~\hbar\omega_{0}/k_{B}$ and set for the coupling strength $\gamma_{s}$ the value $0.1~\omega_{0}$. Throughout this section the system is assumed to be initially in the localized state $|q_{1}\rangle$ belonging to the left well (see Fig.~\ref{fig1}).\\
\indent In Fig.~\ref{fig5} the time evolution of the population $\rho_{11}$ of the state $|q_{1}\rangle$  is shown at two temperatures and for two values of the cutoff frequency $\omega_{c}$.
The time evolutions of $\rho_{11}$ display damped intrawell oscillations ending up in a metastable intrawell equilibrium state which decays further towards the equilibrium over a much larger time scale. The presence of these two different time scales reflects the two different frequency scales of tunneling and intrawell motion in the bare system. Moreover the tunneling dynamics is strongly damped due to the distance between states in different wells. This is because the prefactor $(q_{i}-q_{j})^{2}$ multiplying $\gamma_{s}$ in the influence functional yields a large effective coupling. We also observe that the intrawell oscillations are slower for higher $s$, especially at the higher value of the cutoff frequency. This can be ascribed to a larger renormalized mass (see Eq.~(\ref{M-star})) due to the stronger  presence of high frequency modes, especially for the higher cutoff frequency,  as exemplified by Fig.~\ref{fig2}. Moreover, for both cutoff frequencies, the higher is $s$ the less the oscillations are damped. This is because, on the time scale of the intrawell motion, the bath modes  contributing to the quantum friction are those with $\omega\lesssim\Omega$, which are denser at lower $s$ (see Fig.~\ref{fig2}).\\
\indent Note also that, varying $s$, the long time dynamics has different behaviors for the two cutoff frequencies. In particular, for $\omega_{c}=5~\omega_{0}$ the relaxation is faster at high $s$ while for $\omega_{c}=50~\omega_{0}$ is faster at low $s$.\\ 
\begin{figure}[ht]
\begin{center}
\includegraphics[height=5.7cm,angle=0]{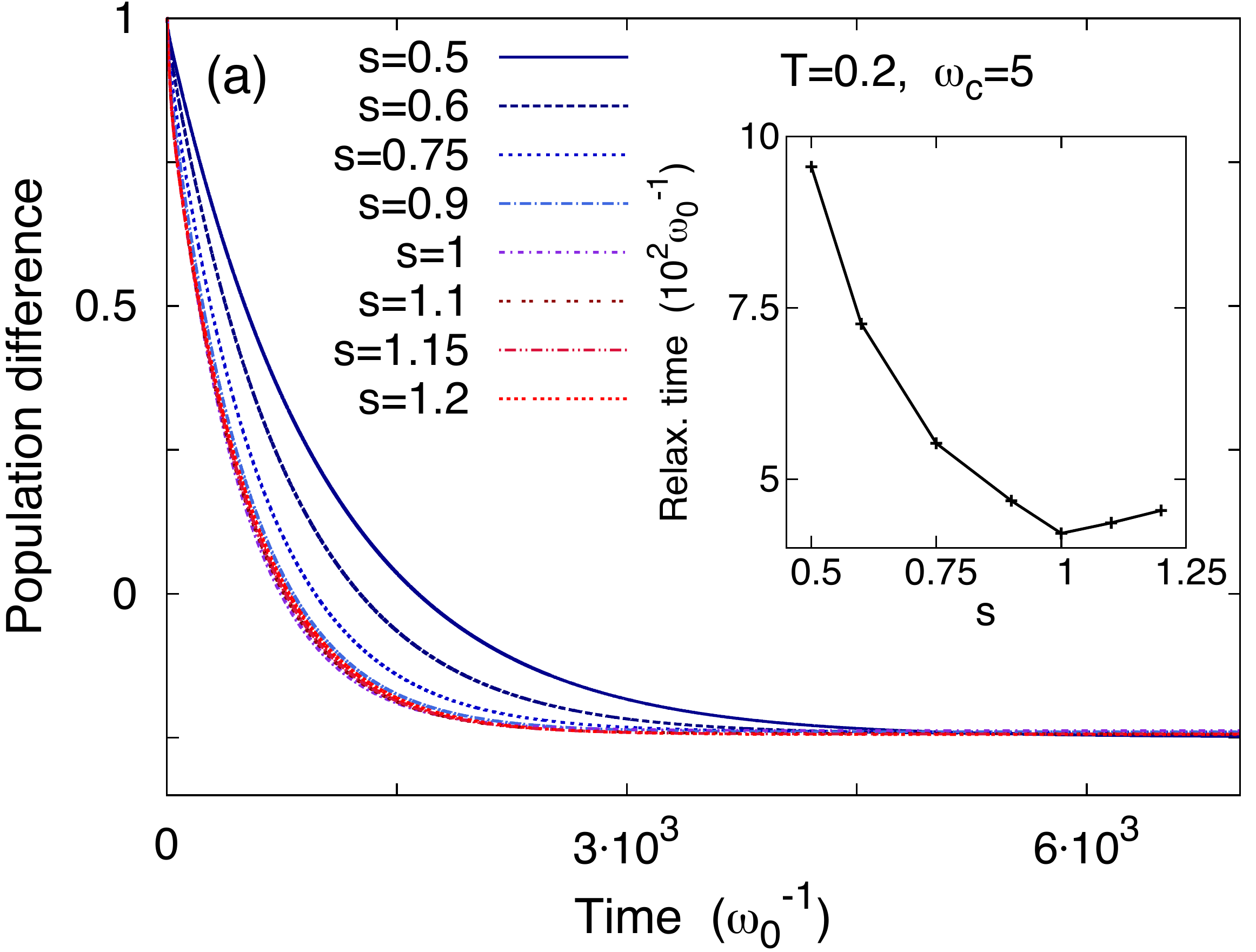}\hspace{0.2cm}
\includegraphics[height=5.7cm,angle=0]{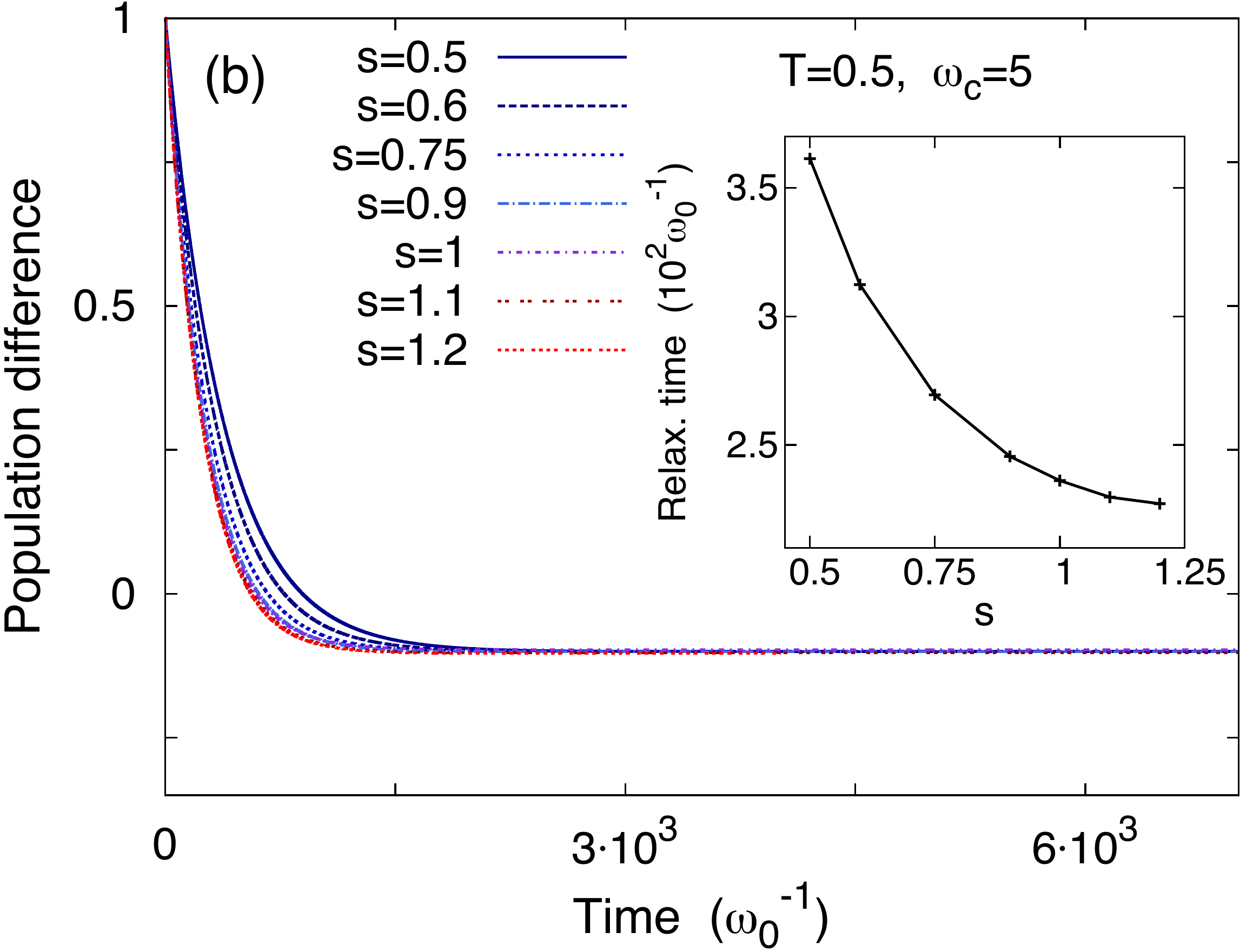}\\
\includegraphics[height=5.7cm,angle=0]{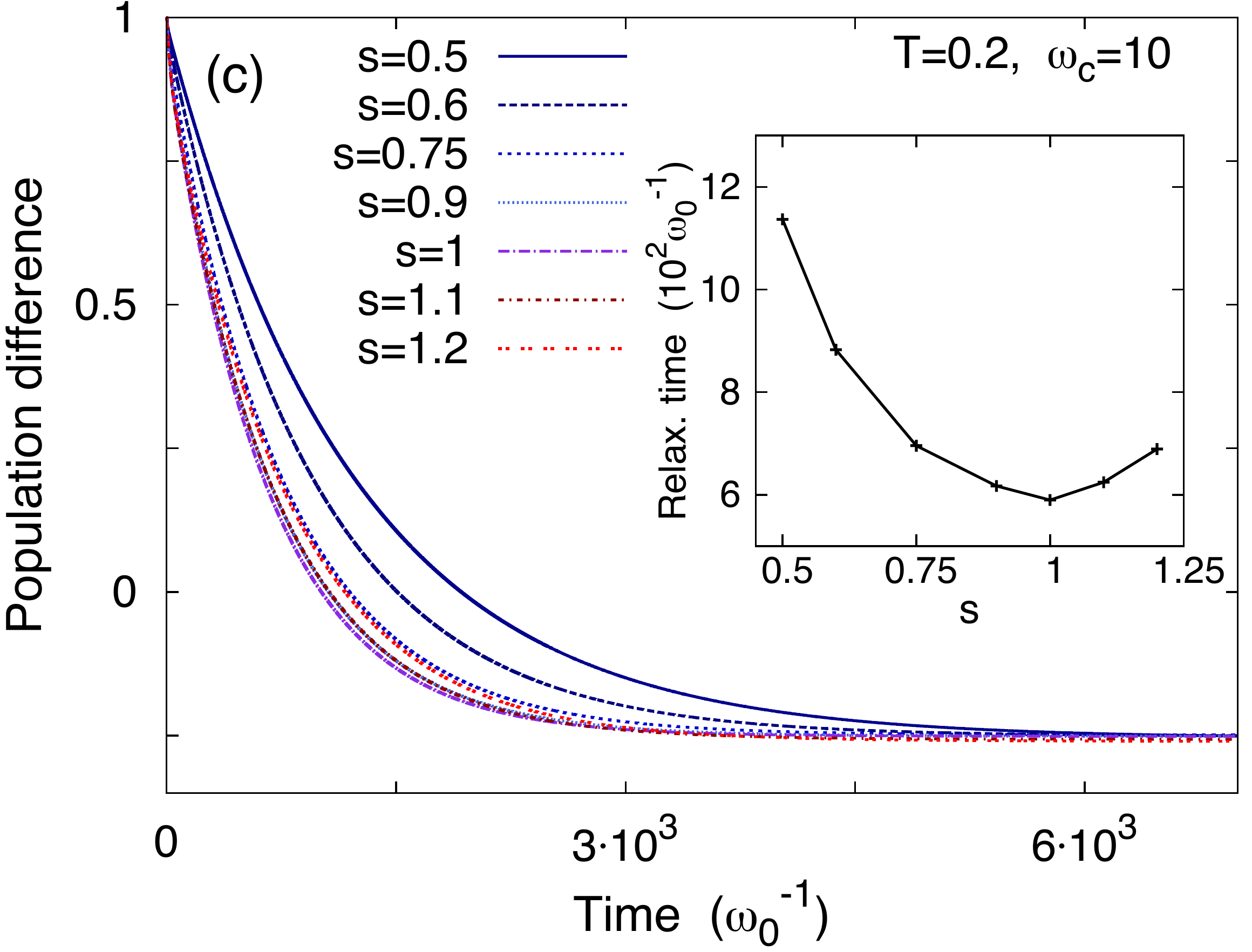}\hspace{0.2cm}
\includegraphics[height=5.7cm,angle=0]{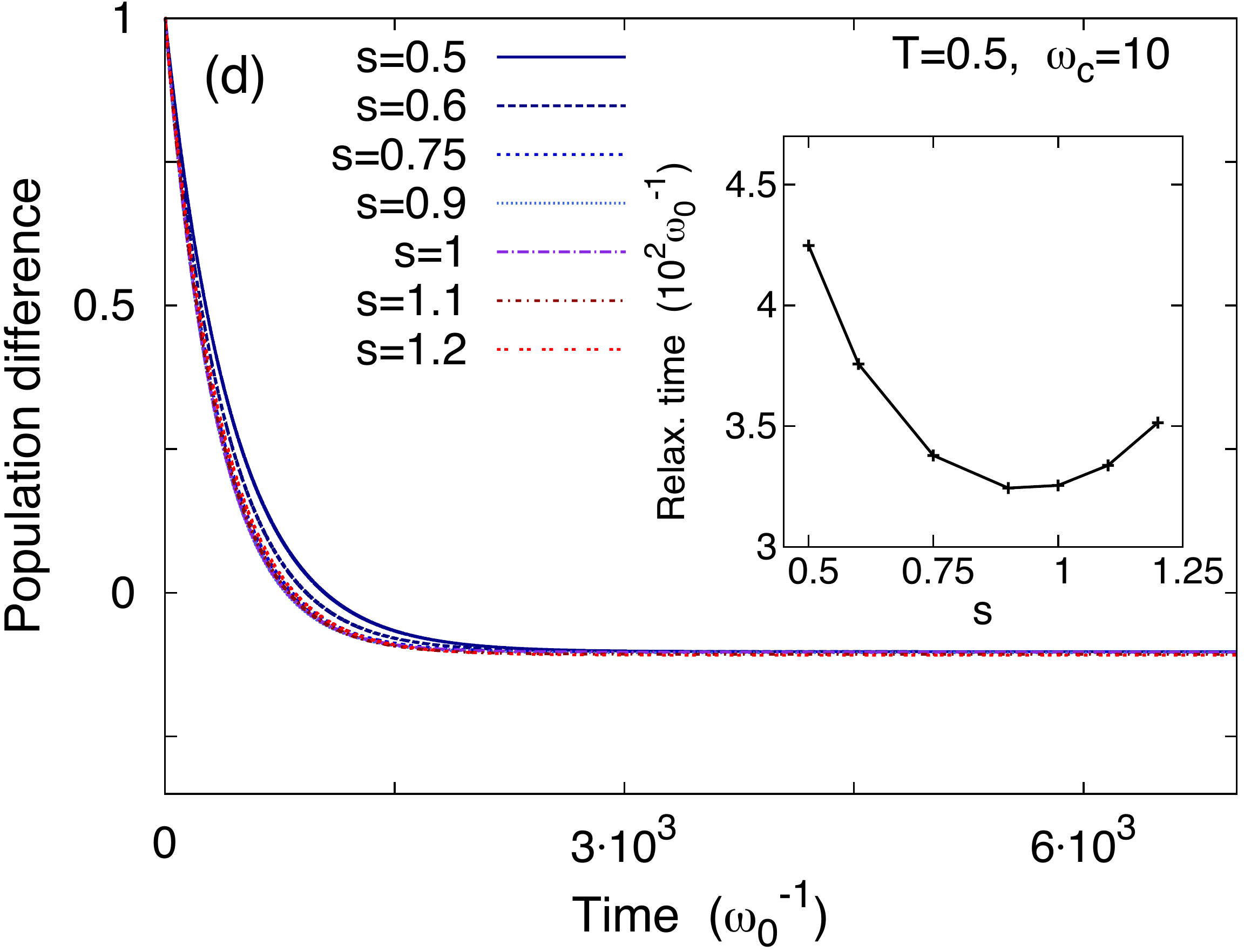}\\
\includegraphics[height=5.7cm,angle=0]{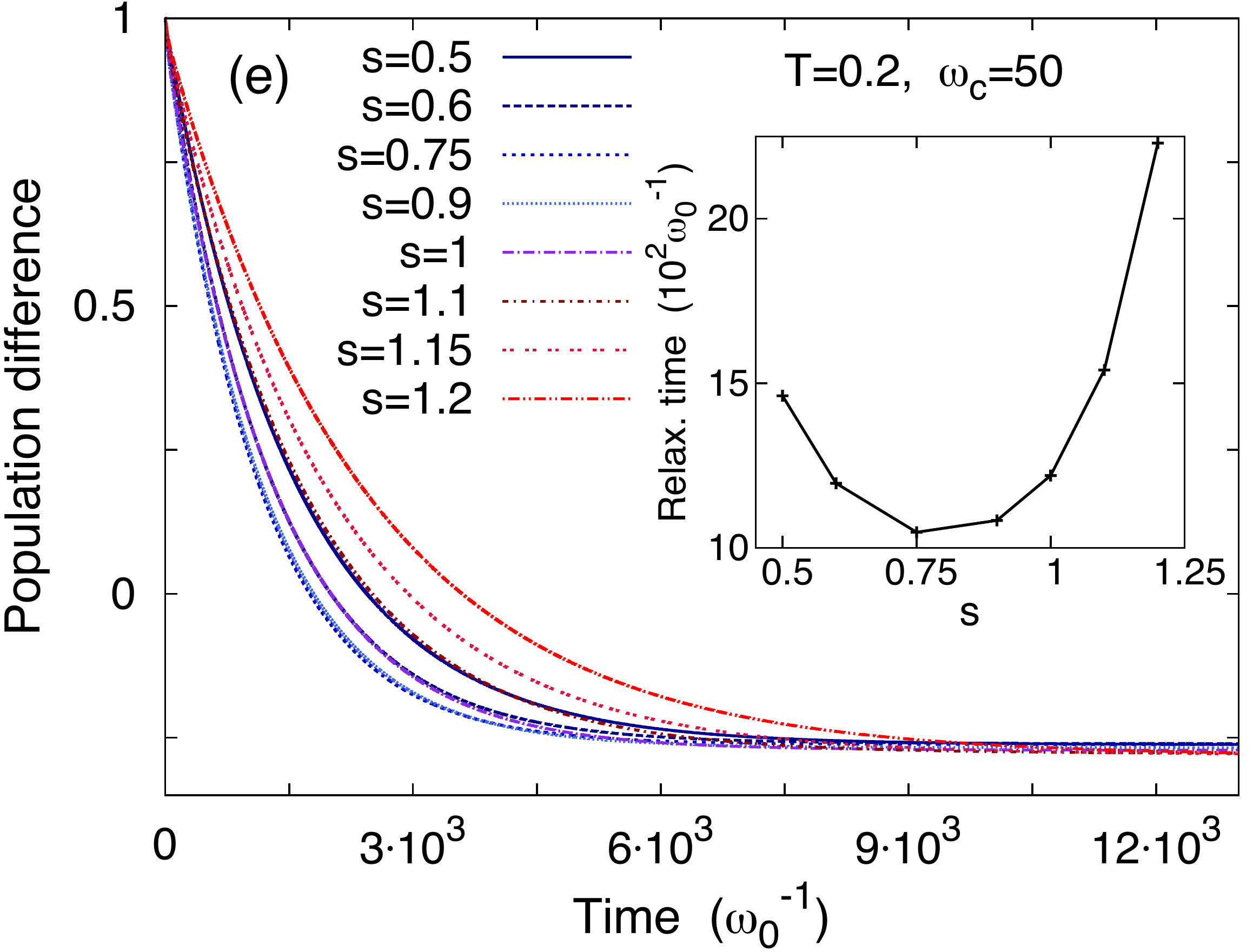}\hspace{0.2cm}
\includegraphics[height=5.7cm,angle=0]{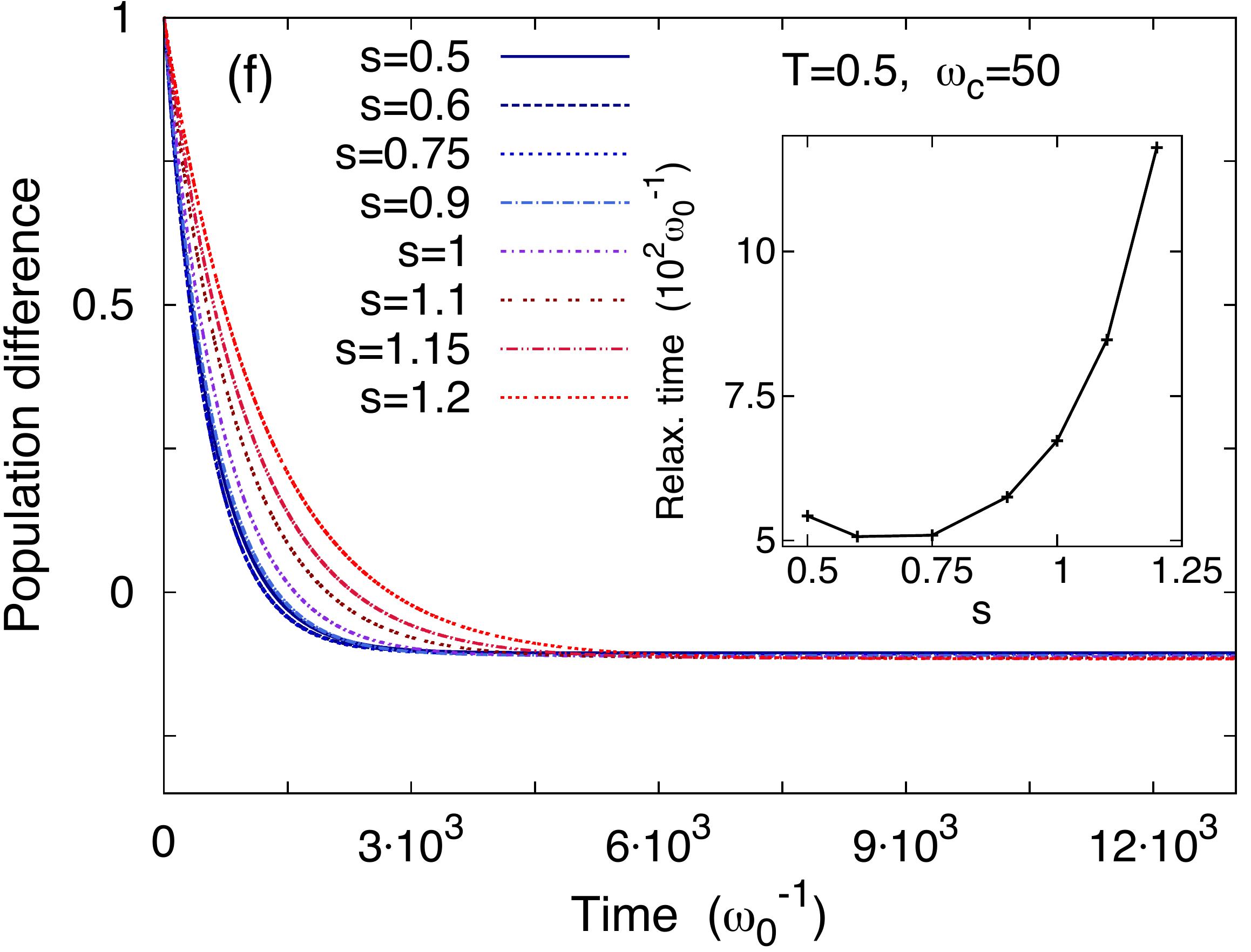}\\
\caption{\small{Time evolution of the population difference $P_{L}-P_{R}$ for different spectral densities ($0.5\ge s\ge 1.2$) at temperatures $T=0.2$ (left panels) and $T=0.5$ (right panels) and at cutoff frequencies $\omega_{c}=5$
(upper panels), $\omega_{c}=50$
(central panels), and  $\omega_{c}=50$ (lower panels). Insets - Relaxation times $\Lambda_{min}^{-1}$ as a function of $s$. The coupling strength is $\gamma_{s}=0.1~\omega_{0}$. Temperatures and frequencies are in units of $\hbar\omega_{0}/k_{B}$ and $\omega_{0}$, respectively.}}
\label{fig6}
\end{center}
\end{figure}
\indent This can be better seen in Fig.~\ref{fig6}, where the time evolution of the population difference $P_{L}-P_{R}$, where $P_{L( R )}=\rho_{11(33)}+\rho_{22(44)}$, is shown for two temperatures and for three values of the cutoff frequency, along with the relaxation time $\Lambda_{\text{min}}^{-1}$ as a function of $s$  (see Section~\ref{GME-DVR} for the definition of  $\Lambda_{\text{min}}$).
The $\omega_{c}$-dependent minima  in the relaxation time as a function of the exponent $s$, shown in the insets of Fig.~\ref{fig6}, emerge as the result of two competing mechanisms. On the one hand, by lowering  the exponent $s$ the density of low-frequency modes  ($\omega \sim \Omega_{1,2}$) of the bath is increased (see Fig.~\ref{fig2}). These modes act on the time scales of the tunneling $\Omega_{1,2}^{-1}$ contributing to the friction exerted by the heat bath. As a consequence, by moving towards smaller values of $s$, the dissipation is enhanced and the tunneling is hampered. On the other hand, the mass renormalization term, defined by Eq.~(\ref{M-star}), increases with $s$ slowing down the relaxation due to the increased inertia of the system. These two competing behaviors yield the minima in the relaxation times. This physical picture is confirmed by the fact that for large $\omega_{c}$, where the mass renormalization effect is stronger, the minimum moves towards lower values of $s$.\\
\begin{figure}[ht]
\begin{center}
\includegraphics[height=5.7cm,angle=0]{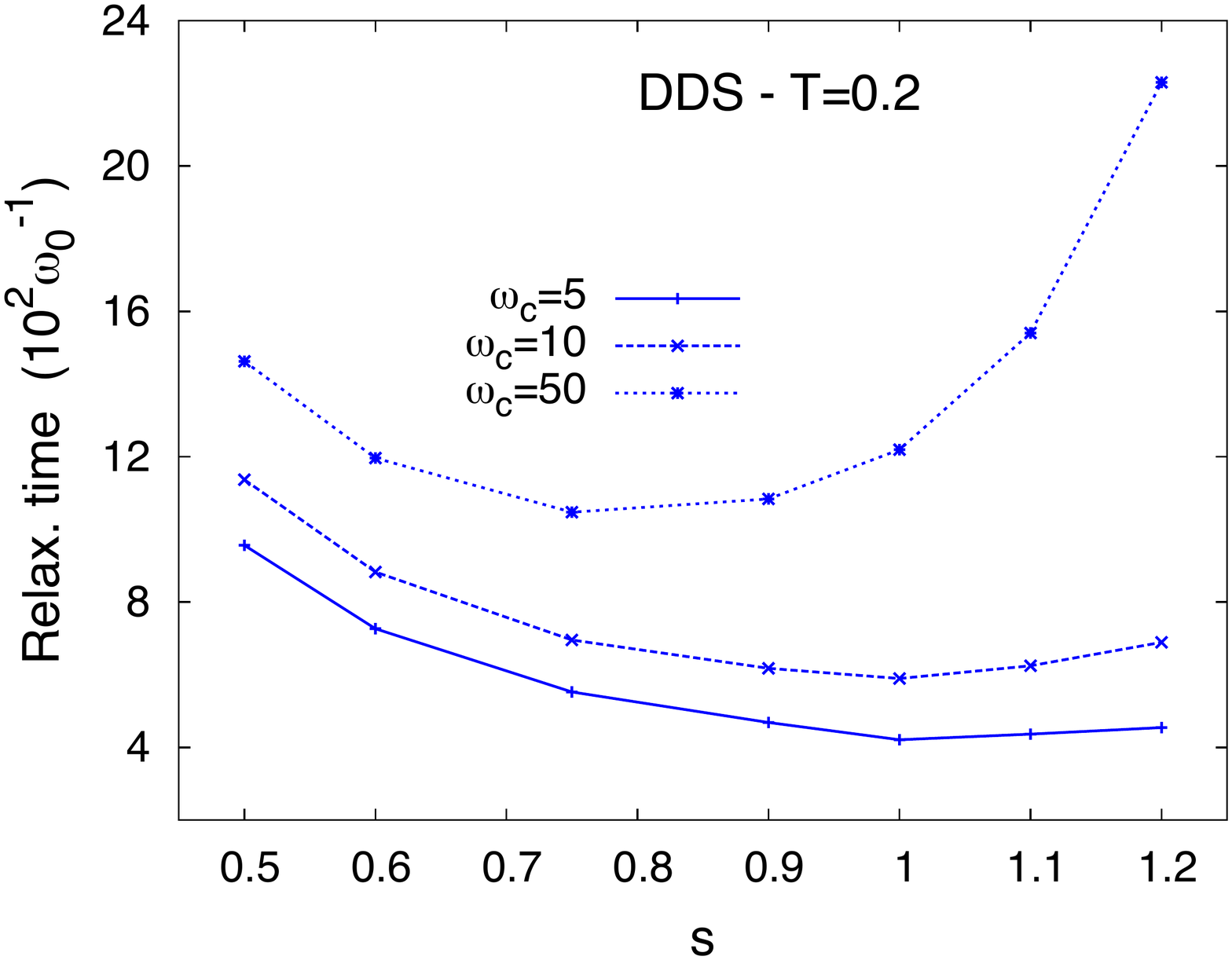}\hspace{0.2cm}
\includegraphics[height=5.7cm,angle=0]{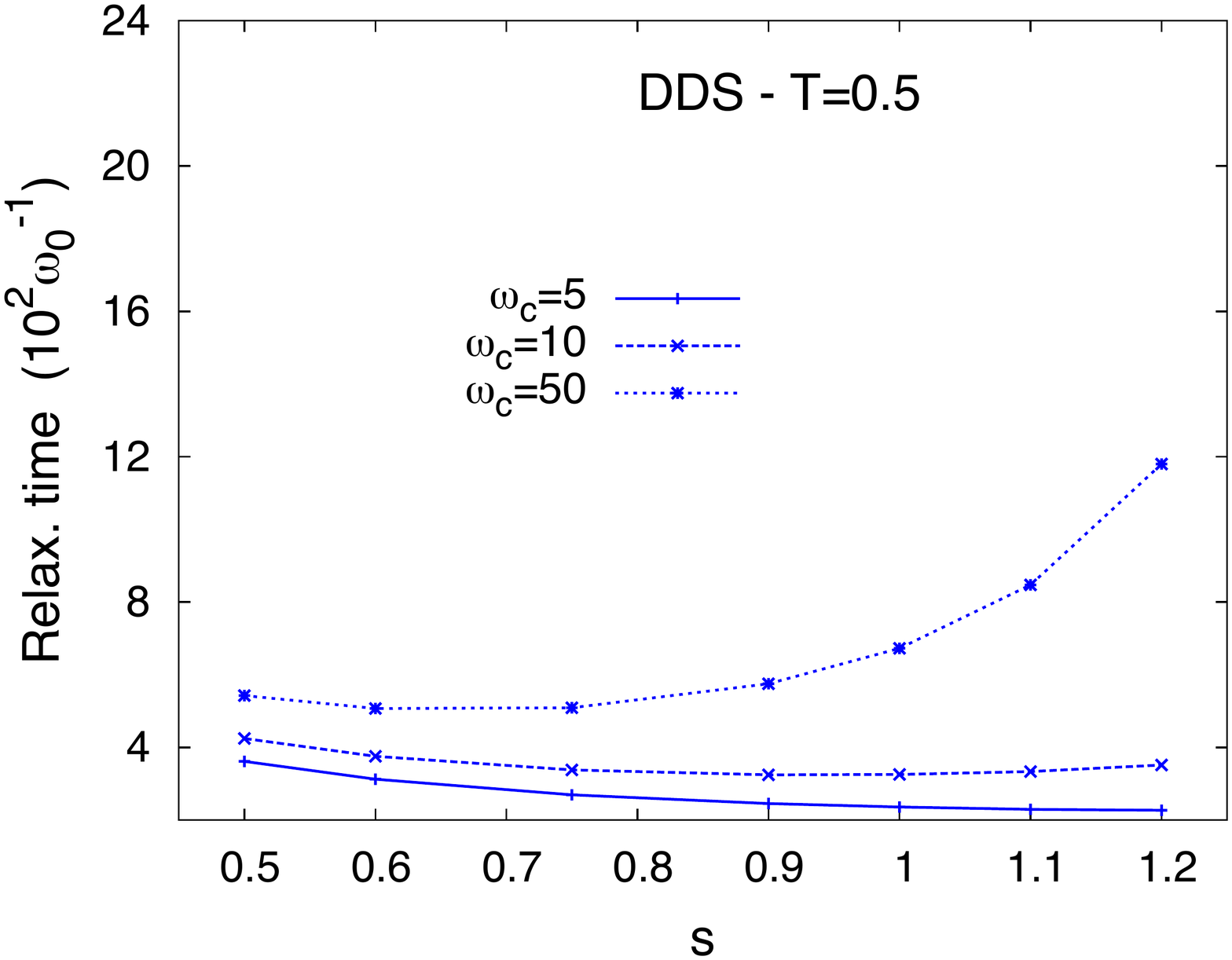}\\
\vspace{0.3cm}
\includegraphics[height=5.7cm,angle=0]{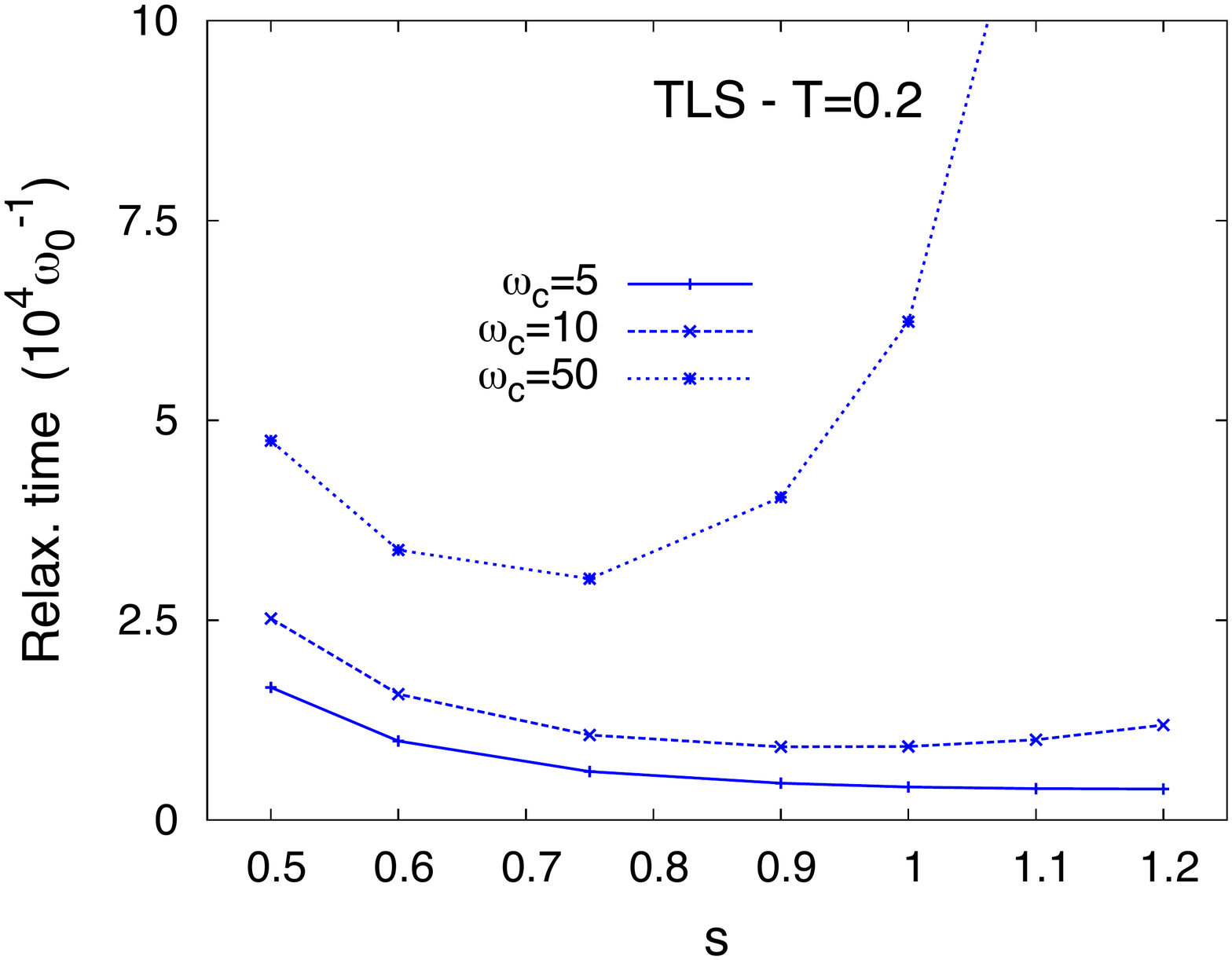}\hspace{0.2cm}
\includegraphics[height=5.7cm,angle=0]{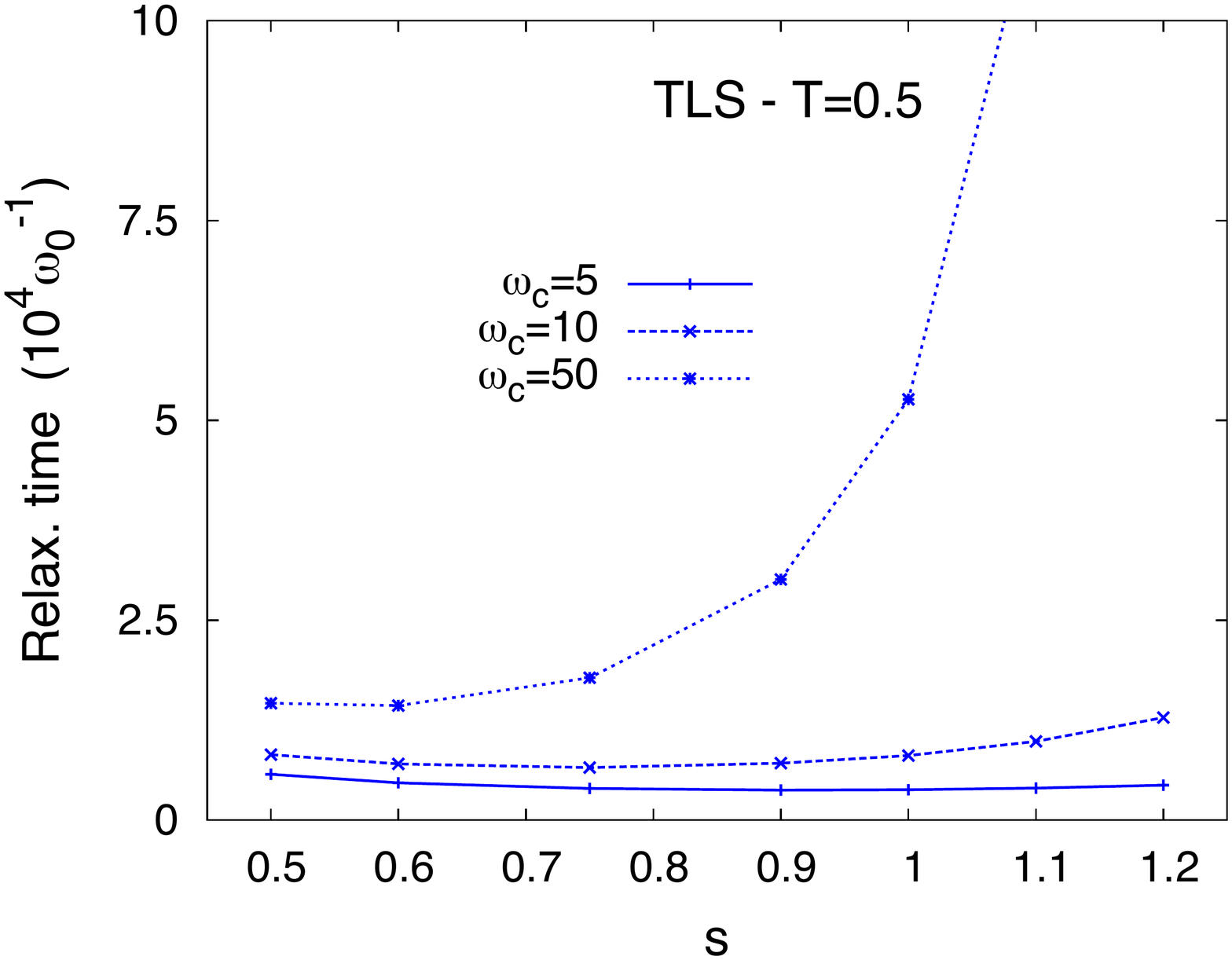}
\caption{\small{Relaxation time vs $s$. Comparison between the four-level system (upper panels, same curves as in the insets of Fig.~\ref{fig6})  and the two-level system (TLS) approximation (lower panels). Temperatures and frequencies are in units of $\hbar\omega_{0}/k_{B}$ and $\omega_{0}$, respectively.}}
\label{fig7}
\end{center}
\end{figure}
\indent To highlight the fact that considering the higher energy doublet yields, in our dissipation regimes, different predictions with respect to a TLS treatment, we calculate the relaxation times vs $s$ for the same physical system within the TLS approximation. The dynamics of $P_{L}$, the population of the left well state, in the incoherent regime is given by the TLS version of Eq.~(\ref{GME-markov}), which has solution $P_{L}(t)=(\Gamma_{LR}+\Gamma_{RL}e^{-\Gamma t})/\Gamma$, where $\Gamma=\Gamma_{LR}+\Gamma_{RL}$. The relaxation time is the inverse of the rate $\Gamma$.\\
\indent The results are shown in the lower panels of Fig.~\ref{fig7}. The relaxation times are almost two orders of magnitude larger in the TLS approximation. This is because the system has a single frequency scale, namely $\Omega_{1}$, and, with respect to this scale, the coupling $\gamma_{s}=0.1~\omega_{0}$ is very strong. Further, the space separation between left state and right state, which coincides with the distance between the potential minima, is large. On the other hand, in the predictions for the double-doublet system, the presence of the higher energy doublet entails the appearance of a second pair of DVR states ($|q_{2}\rangle$ and $|q_{3}\rangle$, see Fig.~\ref{fig1}) which lie closer to each other and allow for less damped tunneling transitions, shortening the relaxation time (upper panels of Fig.~\ref{fig7}).
Nevertheless, the TLS case also shows a minimum.  Notice that, at fixed $s$, the prediction for the TLS is that the relaxation time is not very sensitive to the increase in temperature from $0.2$ to $0.5~\hbar\omega_{0}/k_{B}$. On the contrary, the presence of the higher doublet, which is more excited at higher temperature, causes a speed up of the relaxation for the four-level system by increasing the temperature. This different dynamical behavior, with respect to an increase of $T$, is in agreement with the analysis made in Ref.~\cite{Johnson2011}. There, the presence of an upper energy doublet accounts for the observed enhancement of the tunneling as a function of the temperature. 
\section{Conclusions}
\label{conclusions}
In this work we investigate the multilevel dissipative dynamics of a quantum bistable system strongly interacting with a heat bath of bosonic modes. The study is carried out by using the Feynman-Vernon influence functional approach for the open system dynamics. The dynamics in the dissipation regimes considered is characterized by damped intrawell oscillations and incoherent tunneling. We focus on the influence of the spectral properties of the bosonic heat bath on the transient dynamics. These properties are described by the spectral density function, assumed to be of the form $\omega^{s}$ with a high-frequency cutoff.\\
\indent By varying the exponent $s$, we find that the intrawell oscillations are less damped and slower at higher $s$ and that the relaxation to the equilibrium has a minimum at a value of $s$ which depends on the cutoff frequency. These effects can be accounted for by considering the interplay of the quantum friction exerted by the low frequency part of the bath and the mass renormalization given by the high frequency modes. By comparing the predictions of our multilevel system with those of the TLS version of the same system we find that, in the regime of temperatures considered, the presence of a higher energy doublet cannot be neglected.
\section*{Acknowledgments}  This work was partially supported by MIUR through Grant. No. PON$02\_00355\_3391233$, ÒTecnologie per l'ENERGia e l'Efficienza energETICa - ENERGETICÓ.
\bibliographystyle{iopart-num}
\providecommand{\newblock}{}

\end{document}